\documentclass[twocolumn,showpacs,laps,prl,amsmath,amssymb]{revtex4}

\usepackage{bm,color}
\usepackage{graphicx}
\usepackage{epstopdf}
\usepackage{amsthm}

\usepackage[normalem]{ulem}
\usepackage[usenames,dvipsnames]{xcolor}

\usepackage{lipsum}

\definecolor{ngreen}{rgb}{0.2,0.6,0.2}

\definecolor{golden}{rgb}{0.8,0.6,0.1}

\newcommand{\nn}{\nonumber \\}
\newcommand{\erf}[1]{Eq.~(\ref{#1})}

\newcommand{\ket}[1]{|#1\rangle}
\newcommand{\bra}[1]{\langle #1 |}
\newcommand{\ml}{\zeta}
\newcommand{\xfrac}[2]{{#1}/{#2}}

\newcommand{\PRLsection}\section
\newcommand{\PRLsubsection}\subsection
\newcommand{\PRLnonsection}\section
\newcommand{\sm}[1]{Appendix~{#1}}
\newcommand{\letter}{paper} 

\begin{document}
\title{ Stochastic Heisenberg limit: Optimal estimation of a fluctuating phase}
\author{Dominic W. Berry${}^1$, Michael J. W. Hall${}^2$, and Howard M. Wiseman${}^2$}
\affiliation{${}^1$Department of Physics and Astronomy, Macquarie University, Sydney, NSW 2109, Australia\\
${}^2$Centre for Quantum Computation and Communication Technology (Australian Research Council), Centre for Quantum Dynamics, Griffith University, Brisbane, QLD 4111, Australia}

\begin{abstract}
The ultimate limits to estimating a fluctuating phase imposed on an optical beam can be found using the recently derived continuous quantum Cram\'er-Rao bound.   For  Gaussian stationary statistics, and a phase spectrum scaling asymptotically as $\omega^{-p}$ with $p>1$,  the  minimum mean-square error in any (single-time) phase estimate 
scales as ${\cal N}^{-2(p-1)/(p+1)}$, where ${\cal N}$ is the photon flux. 
This gives the usual  Heisenberg limit for a constant phase (as the limit $p\to\infty$) and  provides a stochastic Heisenberg limit 
for fluctuating phases. For $p=2$ ( Brownian motion), this limit can be attained by phase tracking.
\end{abstract}

\maketitle

Estimating the phase imposed on an optical beam, by nature or by an agent,
is a key task in metrology and communication respectively. 
One case of broad relevance is that where the 
phase varies stochastically in time over a 
wide range  \cite{Science,Berry02,Berry06,erratum,Tsang,Wheatley10,Iwasawa13,Tsang13}.  
It is only very recently that it has been possible 
to experimentally demonstrate the quantum enhancement (by a constant factor) of the estimation of such a strongly fluctuating phase, 
using nonclassical (squeezed) light and homodyne detection with adaptive phase tracking  \cite{Science,Iwasawa13}. 

Adaptive phase tracking is a sophisticated measurement technique whereby the phase of the local oscillator 
(necessary for homodyne detection) is continuously changed in time to follow an 
estimate of the true phase  \cite{Science,Berry02,Berry06,erratum,Tsang,Wheatley10,Iwasawa13}.  
This enables the phase quadrature of the beam to be 
monitored at all times, to a good approximation, maximizing the phase information obtained. 
Previously it has been 
calculated that phase tracking with squeezed light would enable an imposed 
phase to be estimated with a  mean square error (MSE) scaling as ${\cal N}^{-2/3}$ \cite{Berry02,Berry06,erratum}. 
In contrast, for coherent states (no squeezing) only a ${\cal N}^{-1/2}$ scaling can be achieved \cite{Berry02,Berry06,erratum}. 
Here ${\cal N}$ is the mean flux (photons per second) in the beam, and the imposed phase 
is modeled by Brownian motion.

While experiments in optical phase tracking 
have not yet demonstrated  an improvement   
 over the coherent state scaling of ${\cal N}^{-1/2}$,  
the possibility of doing so in the near future raises pressing theoretical questions: 
is the MSE scaling of ${\cal N}^{-2/3}$, derived assuming adaptive estimation \cite{Berry02,Berry06,erratum}, 
the best possible? If not, 
what is the the ultimate limit to estimating a  fluctuating phase  and how can it be achieved?

For measurement of a \emph{constant} phase, the fundamental bound is the Heisenberg limit \cite{heislim,WisMil10}:  
a  phase estimate MSE scaling as $\langle N \rangle^{-2}$, where $\langle N \rangle$ is the mean number
of photons per estimate.
This a quadratic improvement over the  $\langle N \rangle^{-1}$  scaling achievable using coherent 
states (the standard quantum limit,  or SQL)  \cite{heislim,WisMil10}. 
Hence, if quantum mechanics similarly allowed a quadratic improvement in the case of a \emph{fluctuating} phase,
 the corresponding fundamental limit for the MSE 
would scale as ${\cal N}^{-1}$.

Contrary to this intuition, we prove in this \letter, with only weak assumptions, 
that the fundamental bound to estimating  Brownian phase fluctuations is a MSE scaling as 
${\cal N}^{-2/3}$. This establishes that adaptive phase tracking can be a very 
effective measurement technique for this problem, giving an uncertainty  at most  a constant factor 
greater than the minimum allowed by quantum mechanics (under our assumptions).
This ${\cal N}^{-2/3}$ scaling for Brownian fluctuations is just a special case 
of our general stochastic Heisenberg limit, which allows for any inverse power-law describing the phase fluctuation spectrum 
at high frequencies, and which also yields the constant-phase Heisenberg limit as a special case.
 
This \letter\ is organised as follows. First we derive  the  
general stochastic Heisenberg limit, and the stochastic SQL.  
Next we specialize to the scenario 
of Ref.~\cite{Science}: a squeezed beam comprising the output of an optical 
parametric oscillator (OPO) with an added mean field, and a phase 
varying like damped Brownian motion.
 We consider the ultimate limit, and  find the same scaling as in the general case, but with an explicit constant of proportionality, 
consistent with the numerics of Ref.~\cite{erratum}.
 
\PRLsection{General proof} Our result applies to the situation of a continuous beam [a one-dimensional quantum field ${b}(t)$], 
on which there is an imposed phase $\varphi(t)$.
We  require only three conditions: 
\begin{enumerate}
\item  The statistics of the field quadratures and imposed phase  are \emph{stationary}.
\item The statistics of the field quadratures and imposed phase are \emph{Gaussian} and \emph{time-symmetric}. \label{Gaussumption}
\item The phase spectrum scales as $|\omega|^{-p}$ for large $|\omega|$, for some $p>1$. \label{powerassum}
\end{enumerate}
We now explain these conditions in more detail.
The instantaneous creation
operator $b^\dagger(t)$ of the beam obeys $[ b(t), b^\dagger(t')]=\delta(t-t')$,  and 
$\langle  b^\dagger(t)  b(t) \rangle = {\cal N}$ is the photon flux  \cite{WisMil10}.   
The stationarity of its statistics, and those of $\varphi(t)$, means that 
one-time expectation values are constant, and two-time correlations depend only on the time difference. 

The assumed Gaussian statistics mean there is a Gaussian Wigner functional for the
quadratures  \cite{WisMil10}  
\begin{equation}
X(t) :=b^\dagger(t) + b(t), \qquad
Y(t) := i[b^\dagger(t) - b(t)], \label{quads}
\end{equation}
 and a Gaussian distribution for the phase $\varphi(t)$. 
 For a single Gaussian variable, such as  $\varphi(t)$,  the autocorrelation function is automatically time-symmetric. 
However, the  beam  quadratures  $X(t)$ and $Y(t')$  may be correlated, 
and our derivation below requires that this cross-correlation be time-symmetric (i.e. invariant under $t \leftrightarrow t'$).

The third condition allows for a vast range of phase fluctuation models.
For $p=2$ it means that at short times the fluctuations are like Wiener noise  \cite{WisMil10}  (Brownian motion),
which has the spectrum $\kappa/\omega^2$, where $\kappa$ is a constant with units of frequency.
More generally, we take the scaling constant to be $\kappa^{p-1}$, so $\kappa$ still has units of frequency.
In the limit $p\to\infty$ the phase is effectively constant.

In Ref.~\cite{TsangWise}, a continuous form of the quantum Cram\'er-Rao inequality was derived, 
giving a lower bound
on the  MSE   of any unbiased estimate, $\hat\varphi(t)$, of a time-varying parameter $\varphi(t)$, 
\begin{equation} \label{crb}
\langle [\hat \varphi(t)-\varphi(t)]^2\rangle \ge F^{-1}(t,t).
\end{equation}
Here $F(t,t')$ is the Fisher information matrix (with continuous indices $t$ and $t'$) of the phase of the beam, given by a sum of quantum and classical contributions 
\begin{equation} \label{fish}
F(t,t') := F^{(Q)}(t,t') + F^{(C)}(t,t'),
\end{equation}  
and the (matrix) inverse in Eq.~(\ref{crb}) is defined by
\begin{equation} \label{inv}
\int ds  \, F^{-1}(t,s)  F(s,t') = \delta(t-t').
\end{equation} 
For the case where $\varphi(t)$ is an imposed phase, 
\begin{align}
\label{eq:fqdef}
F^{(Q)}(t,t') &:=  4 \left\langle \Delta n(t) \Delta n(t') \right\rangle,    \\
F^{(C)}(t,t')   &:= \int D\varphi \, P[\varphi] \frac{\delta \, \ln P[\varphi]}{\delta \varphi(t)}  \frac{\delta \, \ln P[\varphi]}
{\delta \varphi(t')}. \label{eq:fcdef}
\end{align} 
 In the above, $\int D\varphi \,  \cdots$ denotes an integral over all possible functions $\varphi(t)$, 
the functional $P[\varphi]$ gives the prior weight for each function, and 
$\delta / \delta \varphi(t)$ is a functional derivative. Also, $\Delta n(t)=n(t)-\langle n(t)\rangle$, where 
$n(t) := b^\dagger(t) b(t)$ is the generator of the phase shifts, the photon flux operator.
Because of the stationarity condition,
all quantities dependent on two times $t$ and $t'$ are functions only of $t-t'$.
We will express these quantities explicitly as functions of $t-t'$ from here on. 
In particular, the lower bound in Eq.~(\ref{crb}) will be denoted by $F^{-1}(0)$.

To determine $F^{-1}(0)$, substitute Eq.~(\ref{fish}) into Eq.~(\ref{inv}) and take the Fourier transform, to give
\begin{align}
\label{eq:fcq}
\tilde F^{-1}(\omega) = \frac{1}{\tilde F^{(C)}(\omega)+\tilde F^{(Q)}(\omega) },
\end{align}
for the Fourier transform of $F^{-1}(t-t')$.  The value of $F^{-1}(0)$ is then obtained by integrating this over $\omega$.
Our aim is to determine the minimum possible scaling of this value with the photon flux 
${\cal N}=\langle n(t)\rangle$. 

As we assume the phase fluctuations are Gaussian, we have $F^{(C)}(t-t') = \Sigma^{-1}(t-t')$,
with $ \Sigma(t-t'):=\langle \varphi(t)\varphi(t')\rangle-  \langle \varphi\rangle^2$ \cite{Hall02}. For the case $\tilde\Sigma(\omega)=\kappa^{p-1}/|\omega|^p$,
\begin{equation} \label{finv}
\tilde F^{-1}(\omega) = \frac{\kappa^{p-1}}{|\omega|^p+\kappa^{p-1}\tilde F^{(Q)}(\omega)}.
\end{equation}
More generally, we can obtain the result below with the weaker requirement that the phase spectrum approaches this scaling at high frequencies,
i.e.,  $\tilde\Sigma(\omega)=\Omega( \kappa^{p-1}/|\omega|^p)$  as per  condition \ref{powerassum} 
(see  \sm{D}). 

Next we consider the quantity $F^{(Q)}(t-t')$; this may be simplified to
(see  \sm{A})
\begin{equation} \label{eq:gaussian}
F^{(Q)}(t-t') = 4 {\cal N}\delta(t-t') +
f(t-t')-g(t-t'),
\end{equation}
where, in terms of the quadrature operators (\ref{quads}), 
\begin{align} 
f(t-t') &:=\frac 12\left[ \langle : X(t) X(t'):  \rangle + \langle : Y(t) Y(t'):  \rangle \right]^2 \label{defftt} \\
g(t-t') &:=\langle: X(t) X(t') :\rangle\langle: Y(t) Y(t'):\rangle \nn
& \quad -\frac 12 \left(\langle: X(t) Y(t') :\rangle^2 + \langle: Y(t) X(t'):\rangle^2\right) \nn
& \quad +\frac 12 \left( \langle X\rangle^2 +\langle Y\rangle^2\right)^2. \label{defgtt}
\end{align}
Thus, from Eq.~(\ref{finv}), we obtain
\begin{equation}
\label{eq:lower}
\tilde F^{-1}(\omega)  = \frac {\kappa^{p-1}}{|\omega|^p+ \kappa^{p-1}\left[4{\cal N}+ \tilde f(\omega)-\tilde g(\omega)\right]}.
\end{equation}

The photon flux can be written as
\begin{equation}
{\cal N} = \frac 14 \left(  \langle :X(t)X(t):  \rangle + \langle :Y(t)Y(t):  \rangle \right) ,
\end{equation}
and therefore $f(0)=8{\cal N}^2$. In addition,  using  a  spectral uncertainty principle  and the assumption of time-symmetric correlations, 
it can be shown that ${\cal N}/4  \ge - \tilde g(\omega)$ and $\tilde f(\omega)\ge 0$  (see  \sm{B}). 
Since it is easily shown that $F^{(Q)}(t-t')$ is a positive-definite function, by 
Bochner's theorem $\tilde F^{(Q)}(\omega)\ge 0$ \cite{pos}, and thus the denominator in Eq.\ \eqref{eq:lower} is positive. 
Consequently, replacing $-\tilde g(\omega)$ with ${\cal N}/4$ can only decrease the right-hand side; that is,  with $\ml=17/4$,  
\begin{equation}
\label{eq:lower2}
\tilde F^{-1}(\omega)  \ge \frac {\kappa^{p-1}}{|\omega|^p+ \kappa^{p-1}\left[\zeta{\cal N}+ \tilde f(\omega)\right]}.
\end{equation}

Next,  from  the fact that $f(0)=8{\cal N}^2$, the integral of $\tilde f(\omega)$ is ${\cal I}=16\pi{\cal N}^2$.
This means that $\tilde f(\omega)$ cannot be larger than $\mu$ over a range greater than ${\cal I}/\mu$.
To place a lower bound on $F^{-1}(0)$, when integrating Eq.~\eqref{eq:lower2} we may first omit the range of integration   
where $\tilde f(\omega)>\mu$,
and replace $\tilde f(\omega)$ by $\mu$ over the remaining portion.
Second, we can assume that the range of integration omitted is for the smallest values of $\omega$
because that can only further reduce the value of the integral.
Since this range can be at most ${\cal I}/\mu$ in length, this yields
\begin{align}
\label{eq:bound1}
 F^{-1}(0)  &\ge \frac 1{\pi} \int_{{\cal I}/2\mu}^\infty \frac {\kappa^{p-1} \, d\omega}{\omega^p + \kappa^{p-1}\left (\ml{\cal N}+\mu \right)} \nn
&\ge \frac 1{2\pi} \int_0^\infty \frac {\kappa^{p-1} \, d\omega}{\omega^p + \kappa^{p-1}\left (\ml{\cal N}+\mu \right)} \nn
&\ge \frac {\kappa^{p-1}}{2\pi[\kappa^{p-1}(\ml{\cal N}+\mu)]^{1-1/p}},
\end{align}
where the inequality on the second line holds for $({\cal I}/\mu)^p \le \kappa^{p-1}\left (\ml{\cal N}+\mu \right)$ (see  \sm{C}).

To obtain the strongest lower bound on $F^{-1}(0)$ we consider 
the smallest value of $\mu$ that we can take such that $({\cal I}/\mu)^p \le \kappa^{p-1}\left (\ml{\cal N}+\mu \right)$.
This value is $\mu=\Theta({\cal N}({\cal N}/\kappa)^{(p-1)/(p+1)})$.
We consider scaling for large ${\cal N}/\kappa$, in which case $\mu \gg {\cal N}$, and $\zeta{\cal N}$ 
can be ignored.
Equations \eqref{crb} and \eqref{eq:bound1} thus yield our main result, the lower bound 
scaling for the MSE, 
\begin{equation}
\label{eq:bound2}
\langle [\hat \varphi(t)-\varphi(t)]^2\rangle = \Omega\left( \left( \xfrac{\kappa}{{\cal N}}\right)^{2(p-1)/(p+1)} \right).
\end{equation}
Note this scaling cannot be achieved by a coherent-state beam, for which  $F^{(Q)}(t-t') = 4{\cal N}\delta(t-t')$.   It is easy to show, for this case, by taking  $f=g=0$  in Eq.~(\ref{eq:lower}), that
\begin{equation} \label{eq:bound-coh}
\langle [\hat \varphi(t)-\varphi(t)]^2\rangle_{\rm  SQL} = \Omega\left( \left( \xfrac{\kappa}{{\cal N}}\right)^{(p-1)/p} \right).
\end{equation}
 which we call the stochastic SQL scaling.

In the case of Wiener phase fluctuations ($p=2$), the  stochastic Heisenberg  scaling is $(\kappa/{\cal N})^{2/3}$.
A simplified analysis in Ref.~\cite{Berry02} found that adaptive homodyne measurements can yield this scaling,  but 
it 
took into account neither the photon flux due to the squeezing 
nor the information in the photocurrent noise. 
A more complete analysis, taking both of these terms into account, was performed in Ref.~\cite{erratum}
(correcting an error in the analysis of Ref.~\cite{Berry06}). This verified this scaling of $(\kappa/{\cal N})^{2/3}$ for the MSE. 
That is, the lower bound in Eq.~(\ref{eq:bound2}) is attainable by adaptive measurements for $p=2$.

The limit $p\to\infty$ gives a very slowly varying phase.   
In that case 
\erf{eq:bound2} gives the expected  constant-phase  Heisenberg limit scaling $\langle [\hat \varphi(t)-\varphi(t)]^2\rangle = \Omega(\langle {\cal N} \rangle^{-2})$.
Similarly, \erf{eq:bound-coh} gives the expected  SQL  scaling 
$\langle [\hat \varphi(t)-\varphi(t)]^2\rangle_{\rm  SQL} = \Omega(\langle {\cal N} \rangle^{-1})$.
For all other $p$, the quantum enhancement is less than quadratic, 
and as $p\to 1$ there is no quantum advantage.

\PRLsection{OPO squeezing and OU fluctuations}
Next we specialise to the model of squeezing used in Refs.~\cite{Science,Berry02,Berry06}---a coherent field of real amplitude $\alpha$ added to an Optical Parametric Oscillator (OPO) output---and to phase fluctuations modeled by
Ornstein-Uhlenbeck (OU) noise, with $\tilde\Sigma(\omega)=\kappa/(\lambda^2+\omega^2)$ as in Ref.~\cite{Science}.
Asymptotically this is identical to the Wiener phase spectrum $(p=2)$ analysed above. 
For this beam we have $\langle X \rangle=2\alpha$, $\langle Y \rangle=0$, so Eq.~\eqref{eq:gaussian} becomes
\begin{align}
 F^{(Q)}(t-t') &= 4 {\cal N}\delta(t-t') +4\alpha^2 T_+(t-t') \nn &\quad
+ \left\{ [T_+(t-t')]^2 + [T_-(t-t')]^2 \right\} /2, \label{FQ-opo}
\end{align}
where  $T_\pm(t-t')$ are the normally ordered  correlation  functions for the quadrature fluctuations \cite{Science,c17}
\begin{equation}
T_\pm(t-t') = \langle :\Delta Q_\pm(t) \Delta Q_\pm(t') : \rangle,
\end{equation}
where $Q_+:=X$ and $Q_-:=Y$.
They are given by 
\begin{equation}
\label{eq:scicor}
        T_{\pm}( t-t')  = (R_{\pm}-1)
                      \frac{(1\mp x)\gamma}{4}
                       e^{-(1\mp x)\gamma |t-t'|/2}.
\end{equation}
In Eq.~\eqref{eq:scicor}, $R_\pm$ are the antisqueezing and squeezing levels, respectively, at the center frequency.
For an OPO,  $\gamma$ is the cavity's decay rate \cite{c18} and
$x  \in[0,1)$ is the normalized pump amplitude.  In terms of these quantities, the total photon flux is 
\begin{equation}
\label{eq:flux}
{\cal N} = \alpha^2 + \frac{\gamma}{16} [(R_+-1)(1-x) + (R_--1)(1+x)].
\end{equation}

Substituting these expressions in Eq.~(\ref{FQ-opo}) and taking the Fourier transform yields
\begin{align}
&\tilde F^{(Q)}(\omega) = 4{\cal N}+ 4\alpha^2 (R_+-1) \frac{(1-x)^2\gamma^2}{(1-x)^2\gamma^2+4\omega^2} \nn
&+ \frac {\gamma^3}{16} \left[\frac{(R_+-1)^2(1-x)^3}{(1-x)^2\gamma^2+\omega^2}
+\frac{(R_--1)^2(1+x)^3}{(1+x)^2\gamma^2+\omega^2} \right]. \label{tFQ}
\end{align}
For a coherent state ($R_+=R_-=1$), we would just have $\tilde F^{(Q)}(\omega) = 4{\cal N}$, and we would obtain 
$\kappa/(2 \sqrt{4 {\cal N} \kappa + \lambda^2})$ as the lower bound on the MSE. 
This coherent state limit was first derived by Tsang \textit{et al.}\ (see Eq.~(4.5) in Ref.~\cite{Tsang})  
and scales asymptotically as $(\kappa/{\cal N})^{1/2}$ as expected \cite{Berry02}. 

\PRLsubsection{Comparison with the Science experiment}
 It  seems impossible  to obtain an exact analytical solution for $F^{-1}(0)$ in the case of general  OPO  squeezing. 
However, a useful approximation is to just include the terms in \erf{tFQ} that represent information available from the mean field;  
that is, those terms proportional to $\alpha^2$.  
As in the theory of Ref.~\cite{Berry02},  
the estimation performed in Ref.~\cite{Science} only used the signal from the mean field, so this approximation is relevant to those works.
 As in those works we also  express our results in terms of  $\alpha^2$, rather than ${\cal N}$.  
Then we find 
\begin{equation}
\label{eq:twotermsol}
F^{-1}(0) = \frac{(\kappa/2)\left(1+g^2/\sqrt{\Xi_+ \Xi_-}\right)}{\sqrt{\Xi_+}+\sqrt{\Xi_-}},
\end{equation}
where
\begin{equation}
\Xi_{\pm} = \frac 12 \left( 4\alpha^2\kappa+\lambda^2+g^2 \pm \sqrt{(4\alpha^2 \kappa+\lambda^2-g^2)^2-4d} \right),
\end{equation}
with $g=(1-x)\gamma/2$ and $d=4\kappa\alpha^2(R_+-1)g^2$.

With further simplification this bound on the MSE is comparable to the
results given for adaptive measurements on squeezed states in Ref.~\cite{Science}.
First we note that a mixed squeezed state described by  $R_\pm$, $\gamma$  and $x$  can be assumed to be a combination of a pure squeezed state and classical amplitude noise  (see  \sm{E}), 
where the pure squeezed state is described by $R_-^Q=R_-$, $R_+^Q = 1/R_-$, $x^Q=(\sqrt{R_+}-1)/(\sqrt{R_+}+1)$,
and $\gamma^Q= \gamma (1+x)/(1+x^Q)$.
Then one can determine $F^{(Q)}$ from the parameters for the pure squeezed state.

Using this approach, 
 in  the limit of large bandwidth, $\gamma \to \infty$,
 we  obtain
$\kappa/(2 \sqrt{4 \alpha^2 R_- \kappa + \lambda^2})$ as the lower bound on the MSE. 
 This expression is that shown as trace (iii)  in Fig.~3 of Ref.~\cite{Science}, 
derived from Eq.~(3) of Ref.~\cite{Science}  by taking $\sigma_f\to 0$ (the limit of 
perfectly accurate feedback). 
 Note that this is significantly  below  what was observed in the experiment, because 
the mean-field adaptive algorithm used in the experiment was far from
being perfectly accurate.

\PRLsubsection{Ultimate limit for OPO squeezing and $p=2$}
We have already shown that the lower bound to the MSE  for $p=2$  scales as $(\kappa/{\cal N})^{2/3}$.
Now we show how to determine the \emph{constant} of the scaling for this model assuming 
ideal OPO squeezing with $R_-=1/R_+$. 
We include  all terms  in Eq.~\eqref{tFQ} and 
 express our results in terms of ${\cal N}$.
We  introduce dimensionless (starred) parameters via 
${\cal N} = \kappa {\cal N}_\star$,
$\alpha^2 = \alpha_\star^2 \kappa{\cal N}_\star$,
$\gamma = \gamma_\star \kappa {\cal N}_\star^{5/6}$,
$R_+ = R_\star {\cal N}_\star^{1/3}$, and
$\omega = \omega_\star \kappa {\cal N}_\star^{2/3}$,
where $\gamma_\star>0$ and $R_\star>0$.
Then we obtain in the limit ${\cal N}_\star \to \infty$ (see  \sm{F})
\begin{equation}
\label{eq:sqlim}
F^{-1}(0) = {{\cal N}_\star}^{-2/3} \frac 1{2\pi} \int_{-\infty}^{\infty} \frac{d\omega_\star}{\omega_\star^2+\tilde F_\star^{(Q)}(\omega_\star)},
\end{equation}
with
\begin{align}
\label{eq:fqstar}
\tilde F_\star^{(Q)}(\omega_\star) &=   \frac{4 \gamma_\star^2\alpha_\star^2}{\gamma_\star^2/R_\star+\omega_\star^2} +   \frac{\gamma_\star^3\sqrt{R_\star}/2 }{4\gamma_\star^2/R_\star+\omega_\star^2} .
\end{align}

\begin{figure}[!t]
\centering
\includegraphics[width=0.47\textwidth]{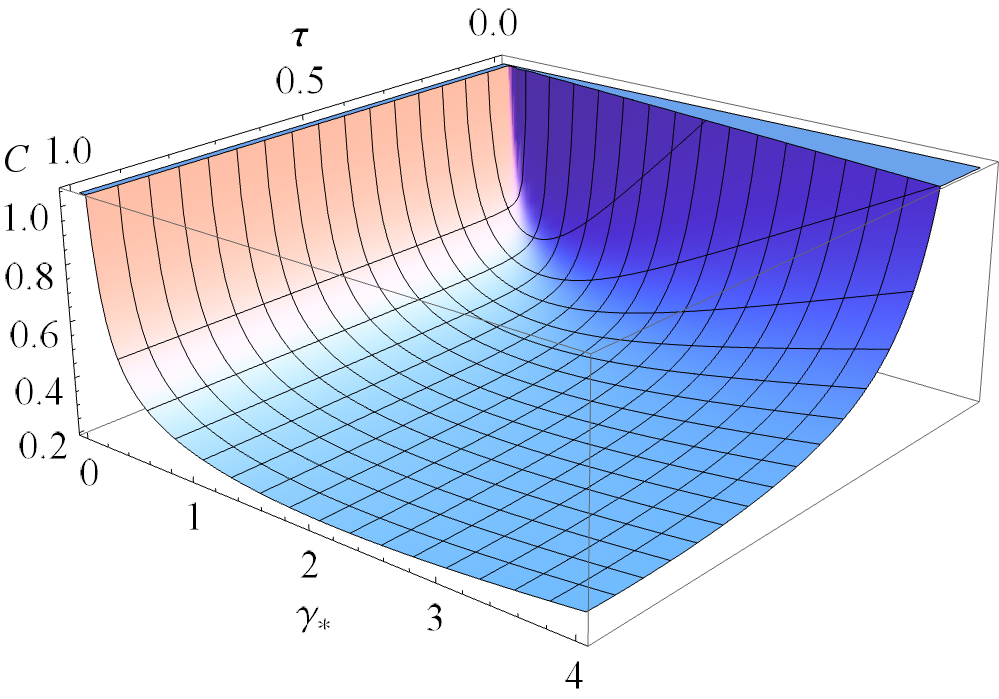}
\caption{ Plot  of $C$  as a function of $\gamma_\star$ and $\tau$ (all dimensionless). }
\label{fig:numerical}
\end{figure}

From Eq.~\eqref{eq:flux}, the above dimensionless parameters are related by $\alpha_\star^2=1-\gamma_\star\sqrt{R_\star}/8$.
It is convenient to define $\tau=\gamma_\star\sqrt{R_\star}/8$, so the allowable values for $\tau$ range from $0$ to $1$.
The value of 
 $C := {{\cal N}_\star}^{2/3}F^{-1}(0)$  was calculated for this range of $\tau$, and $\gamma_\star\in[0,4]$;
the results are given in Fig.~\ref{fig:numerical}.
It can be seen that 
 $C$  is smallest for $\tau=1$, which corresponds to a squeezed vacuum, 
 and the minimum is  
$C_{ 0}=(587-143\sqrt{13})^{1/6}/(4\sqrt{6})\approx 0.20788$ 
for $\gamma_\star=2[2(\sqrt{13}-3)]^{1/3} \approx 2.1319$.
That  is, 
\begin{equation}
\langle [\hat \varphi(t)-\varphi(t)]^2\rangle \gtrsim C_{ 0}  \left( \kappa / {\cal N} \right)^{2/3}, 
\end{equation}
with $C_{ 0}\approx 0.20788$.
In this limit of a squeezed vacuum it is only possible to obtain the estimate of the phase modulo $\pi$.
However, the shallowness of the plot with $\tau$ shows that one can obtain close to the optimal value for
large coherent amplitude, so the phase can be measured modulo $2\pi$.
The  phase tracking  simulations in Ref.~\cite{erratum} 
showed that it is possible to  estimate $\varphi$ modulo $2\pi$ 
with $\langle[\hat \varphi(t)-\varphi(t)]^2\rangle \approx \left( \kappa/ {\cal N} \right)^{2/3}$.  
Moreover those simulations 
 obtained $\hat \varphi(t)$ by filtering the data prior to $t$.  
By using smoothing \cite{Tsang09} of the data before and after $t$, 
one would  halve this MSE \cite{Science,Tsang,Wheatley10,Iwasawa13}. 

\PRLnonsection{Conclusion}

In summary, we have found 
a stochastic form of the Heisenberg limit for 
measurements of a fluctuating phase imposed on a
beam with time-invariant statistics.
For Wiener fluctuations, the scaling of $(\kappa/{\cal N})^{2/3}$ is tight, in that there is a known adaptive measurement scheme that achieves it.
Our bound also reproduces the (tight) Heisenberg scaling of $(\kappa/{\cal N})^{-1}$ for an effectively constant phase. 
We thus  conjecture our general bound to be  tight for all power-law phase spectra.
We do note, however, that we have assumed a beam with  time-symmetric  Gaussian statistics, and it 
is an interesting open question to prove (or disprove) our bound without this assumption.

\acknowledgments{
DWB is funded by an ARC Future Fellowship (FT100100761).  HMW and MJWH are supported by 
 the ARC Centre of Excellence CE110001027.}

\newpage


\onecolumngrid
\newpage {\begin{widetext} \bf \Large Supplemental Material \end{widetext} }
\twocolumngrid

\setcounter{equation}{0}

\renewcommand{\theequation}{S\arabic{equation}}

\section{A: Simplifying the Fisher information for Gaussian states}
Here we show how to obtain Eq.~(9) from (5) in the \letter.
To evaluate averages such as $\langle X^2(t) X^2(t') \rangle$, note that the operators for the different times commute,
so this is the same as a symmetrically ordered moment, and we can use the Wigner function.
This means that the expectation values will satisfy the same rules as for probability distributions.

For Gaussian distributions, one finds that
\begin{equation}
\langle x_1^2 x_2^2 \rangle - \langle x_1^2\rangle \langle x_2^2 \rangle = 2 (\langle x_1 x_2 \rangle^2-\bar x_1^2 \bar x_2^2).
\end{equation}
As a result,
\begin{align}
\label{eq:xrel}
\langle Q_\pm^2(t) Q_\pm^2(t') \rangle &= \langle Q_\pm^2\rangle \langle Q_\pm^2\rangle + 2\langle Q_\pm(t) Q_\pm(t') \rangle^2 \nn
& \quad -2\langle Q_\pm\rangle^2 \langle Q_\pm \rangle^2 ,
\end{align}
where $Q_+:=X$ and $Q_-:=Y$.
Note that for expectation values at a single time, we can omit the time-dependence due to stationarity.
For $t\ne t'$, $X(t)$ and $Y(t')$ commute, so we can again use the Wigner function to take the average, giving
\begin{align}\label{eq:xyrel}
\langle X^2(t) Y^2(t') \rangle &= \langle X^2\rangle  \langle Y^2 \rangle + 2\langle X(t) Y(t') \rangle^2 \nn &\quad 
- 2\langle X\rangle^2 \langle Y \rangle^2 .
\end{align}
On the other hand, to include the case $t=t'$, we need to explicitly symmetrise.
This is not needed for $\langle X^2(t) X^2(t') \rangle$, because it is already symmetric, but it is needed for $\langle X^2(t) Y^2(t') \rangle$.

Using the commutation relation $[X(t), Y(t')] = 2i\delta(t-t')$ enables us to evaluate the symmetrised form
\begin{align}
&\big\langle X^2(t) Y^2(t')\big\rangle_S = \big\langle X^2(t) Y^2(t') \big\rangle \nn & -4i\delta(t-t') \big\langle X(t) Y(t')\big\rangle-2\delta^2(t-t') .
\end{align}
The subscript $S$ indicates the expectation value of the symmetrically ordered operator.
We then obtain
\begin{align}
\langle X^2(t) Y^2(t') \rangle &= \langle X^2(t) Y^2(t') \rangle_S + 4i\delta(t-t')\langle X(t) Y(t')\rangle \nn
& \quad +2 \delta^2(t-t').
\end{align}
Similarly we obtain
\begin{align}
\langle Y^2(t) X^2(t') \rangle &= \langle Y^2(t) X^2(t') \rangle_S - 4i\delta(t-t')\langle Y(t) X(t')\rangle \nn
& \quad+2 \delta^2(t-t').
\end{align}
Then we get
\begin{align}
\langle X^2(t) Y^2(t') \rangle_S &= \langle X^2 \rangle  \langle Y^2 \rangle + 2\langle X(t) Y(t') \rangle_S^2 \nn
& \quad -2\langle X \rangle^2\langle Y \rangle^2 \nn
&=  \langle X^2\rangle  \langle Y^2\rangle + 2\langle X(t) Y(t') \rangle^2 \nn
& \quad - 4i\delta(t-t')\langle X(t) Y(t') \rangle -2\delta^2(t-t') \nn
& \quad -2\langle X \rangle^2\langle Y \rangle^2,
\end{align}
and so the delta functions cancel out to give Eq.~\eqref{eq:xyrel} even for $t=t'$. 
Similarly
\begin{align}
\label{eq:yxrel}
\langle Y^2(t) X^2(t') \rangle &= \langle Y^2 \rangle  \langle X^2\rangle + 2\langle Y(t) X(t') \rangle^2 \nn
& \quad - 2\langle Y\rangle^2 \langle X \rangle^2, 
\end{align}
both for $t=t'$ and for $t\ne t'$.

We can write Eq.~(5) in the form
\begin{align}
 F^{(Q)}(t-t') &= 4 \left[ {\rm Re} \langle b^\dagger(t) b(t) b^\dagger(t') b(t')\rangle \right. \nn
 & \quad \left. - \langle b^\dagger(t) b(t) \rangle \langle b^\dagger(t') b(t')\rangle \right] \nn
 &= \frac 14 \langle [X^2(t)+Y^2(t)][X^2(t')+Y^2(t')]  \rangle \nn & \quad - \frac 14 \langle X^2+Y^2\rangle^2 .
\end{align}
Using the above results for Gaussian states, Eqs.~\eqref{eq:xrel}, \eqref{eq:xyrel}, \eqref{eq:yxrel}, this becomes
\begin{align}
\label{neat}
&F^{(Q)}(t,t') = \frac 12  \left[ \langle X(t) X(t') \rangle^2+\langle Y(t) Y(t')\rangle^2  \right. \nn & \left. + \langle X(t) Y(t') \rangle^2+\langle Y(t) X(t') \rangle^2
- \left(\langle X \rangle^2+\langle Y\rangle^2\right)^2 \right] .
\end{align}
Changing to the normally ordered form we obtain
\begin{align}
\label{eq:expects}
F^{(Q)}(t,t')
&= 4{\cal N}\delta(t-t') -\frac 12 \left( \langle X\rangle^2+\langle Y\rangle^2\right)^2 \nn & \quad +\frac 12  \left[ \langle: X(t) X(t') :\rangle^2+\langle: Y(t) Y(t'):\rangle^2 \right. \nn & \quad \left.+
\langle :X(t) Y(t'): \rangle^2+\langle :Y(t) X(t'):\rangle^2 \right] .
\end{align}
Using the functions $f$ and $g$, Eq.~\eqref{eq:expects} simplifies to Eq.~(9) in the \letter.

\section{B: Bounding $\tilde{g}(\omega)$ and $\tilde{f}(\omega)$}
Let us define the 2-vector $Z(t)$, $2\times2$ Hermitian matrix function $M(t,t')$, and symmetrised correlation function $S_{AB}(t,t')$ by
\begin{align}
Z(t) &:=\left(\begin{array}{c}  X(t)\\ Y(t) \end{array}\right), \nn
M(t,t') &:=\langle Z(t)Z^\dagger(t')+Z(t')Z^\dagger(t) \rangle/2, \nn
S_{AB}(t,t') &:=\langle A(t)B(t')+B(t')A(t)+A(t')B(t)\nn & \qquad +B(t)A(t')\rangle/4.
\end{align}
 Using the commutation relation $[X(t),Y(t')]=2i\delta(t-t')$, one can write $M$ in the form
\begin{equation}
M = \left(\begin{array}{cc}  S_{XX} & S_{XY}+i\openone \\ S_{XY}-i\openone & S_{YY}  \end{array} \right) ,
\end{equation}
where $\openone(t,t'):=\delta(t-t')$.  Note that $M(t,t')$ is positive definite, in the sense that
\begin{equation} \label{posdef}
\int dt\,dt' \,\xi(t) M(t,t') \xi^\dagger(t') = \langle \aleph_1 \aleph_1^\dagger + \aleph_2 \aleph_2^\dagger\rangle/2 \geq 0 
\end{equation}
for any 2-vector function $\xi(t)$, with  $\aleph_1:=\int dt\,\xi(t) Z(t)$ and $\aleph_2:=\int dt\,\xi^\dagger(t) Z(t)$ (this is the continuous analog of the Schr\"odinger-Robertson multimode uncertainty relation \cite{simon}). Stationarity implies further that $M(t,t')$ can be written as $M(t-t')$. Hence, taking Fourier transforms, Eq.~\eqref{posdef} is equivalent to the property
\begin{equation}
\int d\omega \, \tilde{\xi}(\omega) \tilde{M}(\omega) \tilde{\xi}^\dagger(\omega) \geq 0
\end{equation}
for all 2-vector functions $\tilde{\xi}(\omega)$, implying that the $2\times2$ matrix $\tilde{M}$ is positive; i.e., that
\begin{equation} \left( \begin{array}{cc}  \tilde{S}_{XX} & \tilde{S}_{XY}+i \\ \tilde{S}_{YX}-i & \tilde{S}_{YY} \end{array}\right) \geq 0 \end{equation}
(a two-dimensional form of Bochner's theorem \cite{pos}).
This can equivalently be written as the three inequalities $\tilde{S}_{XX}(\omega) \geq 0$, $\tilde{S}_{YY}(\omega) \geq 0$, and
the spectral uncertainty principle
\begin{equation}
\label{eq:genin}
\tilde{S}_{XX}(\omega) \tilde{S}_{YY}(\omega) \ge [\tilde{S}_{XY}(\omega)]^2+1.
\end{equation}
Note that $S_{XX}$, $S_{YY}$, and $S_{XY}$ are time symmetric by construction, and so have real Fourier transforms.

Let us further define
\begin{align}
g^a(t-t') &:= \langle: X(t) X(t') :\rangle\langle: Y(t) Y(t'):\rangle \nn
g^b(t-t') &:=- \frac 12 \left( \langle: X(t) Y(t') :\rangle^2 + \langle: Y(t) X(t'):\rangle^2\right) \nn
g^c(t-t') &:= \frac 12 \left( \langle X\rangle^2+\langle Y\rangle^2\right)^2,
\end{align}
so from Eq.~(11) of the \letter, $g(t-t') =g^a(t-t') +g^b(t-t')+g^c(t-t') $.
Our first aim is to bound $\tilde{g}(\omega)$ in terms of ${\cal N}$ using  the inequality in Eq.~\eqref{eq:genin}.

The inequality (\ref{eq:genin})  means that
\begin{equation}
\label{freqheis}
[1+\tilde h_X(\omega)][1+\tilde h_Y(\omega)]\ge [\tilde h_{XY}(\omega)]^2+1,
\end{equation}
where
\begin{align}
h_X(t-t') &:= \langle: X(t) X(t') :\rangle \nn
h_Y(t-t') &:= \langle: Y(t) Y(t'):\rangle \nn
h_{XY}(t-t') &:= \frac{1}{2}\left[\langle: X(t) Y(t')+ Y(t) X(t'):\rangle \right].
\end{align}
Because $h_X$, $h_Y$, and $h_{XY}$  are time symmetric by construction, $\tilde h_X$, $\tilde h_Y$ and $\tilde h_{XY}$ are real.
Further, the assumption in the \letter\ that the cross-correlation between the quadratures is time symmetric,
i.e., that the symmetrized (or Wigner) cross-correlations satisfy
$\frac{1}{2}\langle X(t)Y(t')+Y(t')X(t)\rangle = \frac{1}{2}\langle X(t')Y(t)+Y(t)X(t')\rangle$,
is equivalent to $\langle :X(t)Y(t'):\rangle = \langle :X(t')Y(t):\rangle$, implying
\begin{equation}
h_{XY}(t-t') = \langle :X(t)Y(t'):\rangle.
\end{equation}

Using the above relations, we can write the photon flux as
\begin{align} \label{Nh}
{\cal N} &= \frac{2}{\pi}\int d\nu \left[ \tilde h_X(\nu) +\tilde h_Y(\nu)\right] \nn
&= \frac{2}{\pi}\int d\nu \left[ \tilde h_X(\omega-\nu) +\tilde h_Y(\omega-\nu)\right].
\end{align}
Similarly 
\begin{align} \label{gah}
\tilde g^a(\omega) &= \frac{1}{2\pi}\int d\nu \, \tilde h_X(\nu) \tilde h_Y(\omega-\nu) \nn
&=  \frac{1}{2\pi}\int d\nu \,  \tilde h_X(\omega-\nu) \tilde h_Y(\nu) ,
\end{align}
and
\begin{equation} \label{gbh}
\tilde g^b(\omega) = - \frac{1}{2\pi}\int d\nu \, \tilde h_{XY}(\nu) \tilde h_{XY}(\omega-\nu) .
\end{equation}
Equations (\ref{Nh}) and (\ref{gah}) then give 
\begin{widetext}
\begin{equation}
\pi{\cal N} + 4\pi\tilde g^a (\omega) = \int d\nu \left\{ [1+\tilde h_Y(\nu)][1+\tilde h_X(\omega-\nu)]+ [1+\tilde h_Y(\omega-\nu)][1+\tilde h_X(\nu)]-2 \right\} .
\end{equation}
Replacing $[1+\tilde h_X]$ by  the smaller expression $[ \tilde h_{XY}^2+1 ]/[1+\tilde h_Y]$ as per \erf{freqheis} gives the inequality 
\begin{equation}
\pi{\cal N} + 4\pi\tilde g^a (\omega) \ge \int d\nu \left\{\frac{[\tilde h_Y(\nu)-\tilde h_Y(\omega-\nu)]^2}{[1+\tilde h_Y(\nu)][1+\tilde h_Y(\omega-\nu)]} 
+[\tilde h_{XY}(\nu)]^2\frac{1+\tilde h_Y(\omega-\nu)}{1+\tilde h_Y(\nu)}+[\tilde h_{XY}(\omega-\nu)]^2\frac{1+\tilde h_Y(\nu)}{1+\tilde h_Y(\omega-\nu)} \right\}.
\end{equation}
\end{widetext}
Each of the three terms here is positive. Dropping the first term and using the inequality $a + b \geq 2\sqrt{ab}$, gives 
\begin{align}
\pi{\cal N} + 4\pi\tilde g^a (\omega)&\ge 2\int d\nu \, |\tilde h_{XY}(\nu) \tilde h_{XY}(\omega-\nu)|.
\end{align}

Because the absolute value cannot exceed the value, it follows that
${\cal N} + 4\tilde g^a (\omega)+ 4\tilde g^b (\omega)\ge 0$.
Moreover, $g^c(t-t')$ is a positive constant by stationarity, and therefore $\tilde g^c(\omega) = 2\pi g^c(0)\delta(\omega) \ge 0$.
This means that ${\cal N} + 4\tilde g (\omega) \ge 0$.

Turning now to $\tilde f(\omega)$, from Eq.~(10) we can similarly write this as 
\begin{equation}
\tilde f(\omega) = \int \! d\nu \left[ \tilde h_X(\nu)+\tilde h_Y(\nu)\right]\left[ \tilde h_X(\omega-\nu)+\tilde h_Y(\omega - \nu)\right] \!.
\end{equation}
Using  
$\tilde h_X(\omega)+\tilde h_Y(\omega)  =  \tilde h_X(\omega) + [1+\tilde h_Y(\omega)] - 1$ 
 and replacing $1+\tilde h_Y(\omega)$ by the smaller term $1/[1+\tilde h_X(\omega)]$,
as per the weaker version of \erf{freqheis} with $1$ on the right-hand-side, and simplifying yields 
\begin{equation}
\tilde h_X(\omega)+\tilde h_Y(\omega)  \ge \frac{[\tilde h_X(\omega)]^2}{1+\tilde h_X(\omega)} \ge 0.
\end{equation}
As $\tilde f(\omega)$ is thus an integral of a product of two non-negative quantities, we immediately get
$\tilde f(\omega)\ge 0$.

\section{C: Details of the inequalities in Eq.~(15)}
The inequality in the second line of Eq.~(15) of the \letter\ is because the integral over the interval $[0,2{\cal I}/\mu]$
cannot be more than half that over the full range, as we now prove. 
The integral over the full range $[0,\infty)$ is given by
\begin{align}
\label{eq:fullint}
&\frac 1{\pi} \int_0^\infty \frac {\kappa^{p-1} \, d\omega}{\omega^p + \kappa^{p-1}\left (\ml{\cal N}+\mu \right)} \nn
&= \frac {\kappa^{p-1}}{\pi[\kappa^{p-1}(\ml{\cal N}+\mu)]^{1-1/p}} \frac 1{{\rm sinc}(\pi/p)} \nn
&\ge \frac {\kappa^{p-1}}{\pi[\kappa^{p-1}(\ml{\cal N}+\mu)]^{1-1/p}}.
\end{align}
In addition, if $({\cal I}/\mu)^p \le \kappa^{p-1}\left (\ml{\cal N}+\mu \right)$, then
\begin{align}
\label{eq:partint}
&\frac 1{\pi} \int_0^{{\cal I}/2\mu} \frac {\kappa^{p-1} \, d\omega}{\omega^p + \kappa^{p-1}\left (\ml{\cal N}+\mu \right)} \nn
&\le \frac 1{\pi} \int_0^{[\kappa^{p-1}\left (\ml{\cal N}+\mu \right)]^{1/p}/2} \frac {\kappa^{p-1}\, d\omega }{\kappa^{p-1}\left (\ml{\cal N}+\mu \right)} \nn
&= \frac {\kappa^{p-1}}{2\pi[\kappa^{p-1}(\ml{\cal N}+\mu)]^{1-1/p}}, 
\end{align}
Comparing this to Eq.~\eqref{eq:fullint} gives the required result. 

\section{D: More general types of phase fluctuations}
\label{app:gen}
We allow for phase fluctuations satisfying the condition
\begin{equation}
\exists \beta:~ \forall |\omega|>\kappa\beta,~ \tilde F^{(C)}(\omega) \le |\omega|^p/\kappa^{p-1}.
\end{equation}
Note that we  continue to  assume that the correlations are  stationary, as per  condition 1. 
For \emph{Gaussian} phase statistics, $\tilde F^{(C)}(\omega) = \tilde \Sigma^{-1}(\omega)$, although more generally
$\tilde F^{(C)}(\omega) \ge \tilde\Sigma^{-1}(\omega)$
 (since $F^{(C)}(t,t') - \Sigma^{-1}(t,t')$ is a positive definite function  \cite{Hall02}). 
Our condition means that, for large $\omega$, $\tilde\Sigma(\omega)$ is greater than or equal to a function scaling as $\kappa^{p-1}/|\omega|^p$.
We could also give the condition directly in terms of $\tilde\Sigma(\omega)$ and it would be equivalent for Gaussian phase statistics, but this form of the condition also allows for non-Gaussian phase statistics in Eq.~(7).
A particular example, for $p=2$ (and Gaussian statistics),  is the case of phase fluctuations modelled by Ornstein-Uhlenbeck noise, where $\tilde \Sigma(\omega)=\kappa/(\lambda^2+\omega^2)$.

Selecting a suitable value of $\beta$, 
then  for $|\omega|>\kappa\beta$,
\begin{equation}
\label{eq:reseq}
\tilde F^{-1}(\omega) \ge \frac{1}{|\omega|^p/\kappa^{p-1}+\tilde F^{(Q)}(\omega)}.
\end{equation}
Following the same derivation as in the body of the \letter, we obtain
\begin{equation}
F^{-1}(0)  \ge \frac 1{\pi} \int_{{\cal I}/2\mu+\kappa\beta}^\infty \frac {\kappa^{p-1} \,d\omega  }{\omega^p + \kappa^{p-1}\left (\ml{\cal N}+\mu \right)}.
\end{equation}
Here, we now omit the smallest range of integration consistent with $|\omega|>\kappa\beta$ (as Eq.~\eqref{eq:reseq}
requires $|\omega|>\kappa\beta$), as well as the range $|\omega|\le \kappa\beta$.
As in the body of the \letter, omitting part of the range of the integral can only further reduce the value of the integral.

As we take $\mu=\Theta({\cal N}({\cal N}/\kappa)^{(p-1)/(p+1)})$ and consider large ${\cal N}/\kappa$,
we obtain ${\cal I}/2\mu>\kappa\beta$, so we again obtain Eq.~(15) from the \letter.
Therefore the more general type of phase fluctuations we have allowed do not alter the scaling. 

\section{E: The Fisher information for mixed states}
\label{app:mixed}
A mixed phase-dependent state can be written in the form
\begin{equation}
\rho(\varphi) = \sum_{\ell} p_{\ell}\rho_{\ell}(\varphi).
\end{equation}
Consider the Fisher information for estimation of the unknown phase shift $\varphi$ via the optimal 
measurement with positive operator-valued measure (POVM) $\{M_k\}$.
We use $q_k$ for the probability of measurement result $k$ on $\rho(\varphi)$.
To upper bound the Fisher information obtainable by this POVM, 
we can always consider a more informative measurement described by 
POVM $\{\ket{\ell}\bra{\ell}\otimes M_k\}$ on the expanded state $\rho(\varphi) = \sum_{\ell} p_{\ell}\ket{\ell}\bra{\ell}\otimes \rho_{\ell}(\varphi)$. 
We use $p_{k\ell}$ to indicate the probability of measurement result $(k,\ell)$ with this expanded POVM, and $p_{k|\ell}=p_{k\ell}/p_\ell$.
Because the expanded POVM cannot result in less information, we must have \cite{Toth}
\begin{align}
F^{(Q)}(\rho(\varphi)) &= \sum_{k} q_{k} \left[\frac d{d\varphi} \log(q_{k}) \right]^2 \nn
&\le  \sum_{k\ell} p_{k\ell} \left[\frac d{d\varphi} \log(p_{k\ell}) \right]^2 \nn
&= \sum_{\ell} p_{\ell}\sum_{k} p_{k|\ell} \left[\frac d{d\varphi} \log(p_{k|\ell}) \right]^2 \nn
& \le \sum_\ell p_\ell F_Q(\rho_\ell(\varphi)).
\end{align}
As a result, the Fisher information is no greater than that for the individual states in the mixture.

When considering a mixture of phase-squeezed states, we have no $XY$ correlations, and therefore require $S_{XX} S_{YY}\ge \openone$.
We are considering the case that the $Y$ quadrature is squeezed, and there is antisqueezing in the $X$ quadrature.
For mixed squeezed states, we can model the $X$ fluctuations as a sum of classical and quantum fluctuations.
That is, we define $S_{XX}^Q:=\openone/S_{YY}$ and $S_{XX}^C := S_{XX} - S_{XX}^Q$, so
$S_{XX} = S_{XX}^Q+S_{XX}^C$.
The inequality $S_{XX} S_{YY}\ge \openone$ means that $S_{XX}^C$ is positive semidefinite,
and therefore represents valid classical fluctuations.

Moreover, it is easily shown that the correlations $S_{XX}^Q$ and $S_{YY}$ correspond to a pure squeezed state.
For this pure squeezed state, we take $R_+^Q = 1/R_-$, and adjust the values of $x$ and $\gamma$ to be consistent.
That is, $x^Q$ and $\gamma^Q$ such that $x^Q$ is determined from $R_+^Q$ and $R_-$,
and $\gamma^Q$ satisfies $\gamma^Q (1+x^Q) = \gamma (1+x)$. 

\section{F: Derivation of Eq.~(25)}
Substituting the equations with the starred quantities into the equation for $F^{-1}(0)$, one obtains
\begin{equation}
\label{eq:sqder}
F^{-1}(0) = {{\cal N}_\star}^{-2/3} \frac 1{2\pi} \int_{-\infty}^{\infty} \frac{d\omega_\star}{\omega_\star^2+[\lambda/\kappa+\tilde F^{(Q)}(\omega)]/\kappa{\cal N}_\star^{4/3}}.
\end{equation}
We use Eq.~(22) of the \letter, and take $R_-=1/R_+$ and $x=(\sqrt{R_+}-1)/(\sqrt{R_+}+1)$.
In the limit of large $R_+$ we have $x=1-2R_+^{-1/2}+O(R_+^{-1})$.
Keeping only the leading order term, we have $1-x\approx 2R_+^{-1/2}$.
Similarly, keeping only leading order terms we have $R_+-1\approx R_+$, $R_- -1 \approx -1$, and $1+x\approx 2$.
Then Eq.~(22) simplifies to
\begin{align}
&\tilde F^{(Q)}(\omega) \approx 4{\cal N}+ 4\alpha^2 R_+ \frac{(4/R_+)\gamma^2}{(4/R_+)\gamma^2+4\omega^2} \nn
&+ \frac {\gamma^3}{16} \left[\frac{8\sqrt{R_+}}{(4/R_+)\gamma^2+\omega^2}
+\frac{8}{4\gamma^2+\omega^2} \right].
\end{align}
For large $R_+$ the second term in the square brackets can be omitted.
Substituting the starred quantities then gives
\begin{align}
\tilde F^{(Q)}(\omega) &\approx 4\kappa{\cal N}_\star+ 4\alpha_\star^2 \kappa {\cal N}_\star^{4/3}  \frac{\gamma_\star^2}{\gamma_\star^2/R_\star+\omega_\star^2} \nn
&\quad + {\gamma_\star^3\kappa{\cal N}_\star^{4/3}} \frac{\sqrt{R_\star}/2}{4\gamma_\star^2/R_\star+\omega_\star^2}.
\end{align}
Now we take $\tilde F_\star^{(Q)}(\omega_\star) :=\tilde F^{(Q)}(\omega)/\kappa{\cal N}_\star^{4/3}$, so that
\begin{equation}
\tilde F^{(Q)}(\omega) \approx 4{\cal N}_\star^{-1/3}+ \frac{4\gamma_\star^2\alpha_\star^2}{\gamma_\star^2/R_\star+\omega_\star^2}  + \frac{\gamma_\star^3\sqrt{R_\star}/2}{4\gamma_\star^2/R_\star+\omega_\star^2}.
\end{equation}
The first term is higher order, and omitting it gives Eq.~(26) of the \letter.
Omitting the term in $\lambda$ and using $\tilde F_\star^{(Q)}(\omega_\star)$ in Eq.~\eqref{eq:sqder} then gives
Eq.~(25) of the \letter.


\begin{thebibliography}{99}
\bibitem{Science} H. Yonezawa  {\em et al.},  
Science \textbf{337}, 1514 (2012).  
\bibitem{Berry02} D. W. Berry and H. M. Wiseman, Phys. Rev. A \textbf{65}, 043803 (2002).
\bibitem{Berry06} D. W. Berry and H. M. Wiseman, Phys. Rev. A \textbf{73}, 063824 (2006).
\bibitem{erratum} D. W. Berry and H. M. Wiseman, Phys. Rev. A \textbf{87}, 019901(E) (2013).
\bibitem{Tsang} M. Tsang, J. H. Shapiro, and S. Lloyd, Phys. Rev. A \textbf{79}, 053843 (2009).
\bibitem{Wheatley10}
   T. A. Wheatley {\em et al.}, 
  {Phys. Rev. Lett.} {\bf 104}, 093601 (2010). 
\bibitem{Iwasawa13}  K. Iwasawa {\em et al.}, eprint arXiv:1305.0066 
\bibitem{Tsang13}  M. Tsang, New J. Phys. \textbf{15}, 073005 (2013).
\bibitem{heislim} M. J.  Holland and K.  Burnett, Phys.  Rev.  Lett.   \textbf{71},  1355 (1993);
 Z. Y. Ou, Phys.  Rev.  A {\bf 55}, 2598 (1997);  D.  Braun and J.  Martin, Nature Commun. {\bf 2}, 223 (2011).  
\bibitem{WisMil10} H. M. Wiseman and G. J. Milburn, \emph{Quantum Measurement and Control}
  (Cambridge University Press, 2010).
\bibitem{TsangWise} M. Tsang, H. M. Wiseman, and C. M. Caves, Phys. Rev. Lett. \textbf{106}, 090401 (2011).
\bibitem{Hall02} M. J. W. Hall, K. Kumar, and M. Reginatto, arXiv:hep-th/0206235 (2002).
\bibitem{pos} R. Bhatia, {\it Positive Definite Matrices} (Princeton University Press, New Jersey, USA, 2007), Sec.~5.5.
\bibitem{Tsang09} M. Tsang, Phys. Rev. Lett. \textbf{102}, 250403 (2009)
\bibitem{c17} {In Ref.~\cite{Science} the correlations, denoted  $X_\pm(\tau)$, are symmetrically ordered so that 
$X_\pm(\tau) = T_\pm(\tau) + \delta(\tau)$.}
\bibitem{c18} {In Ref.~\cite{Science} $2\Delta\Omega_0$ is used instead of $\gamma$.}
\bibitem{simon} R. Simon, N. Mukunda, and B. Dutta, Phys. Rev. A \textbf{49}, 1567 (1994).
\bibitem{Toth} G. T\'{o}th and D. Petz, Phys. Rev. A \textbf{87}, 032324 (2013).
\end{thebibliography}
\end{document}